\documentclass[a4paper]{jpconf}
\usepackage{citesort}

\usepackage{graphicx}
\usepackage{bm}

\def\e#1{\,{\rm e}#1}

\def\vec#1{{\overrightarrow{#1}}}

\def\and{\,\,\&\,}

\def\i{{\bf i}}

\def\skipm#1{}

\def\simge{\mathrel{%
   \rlap{\raise 0.511ex \hbox{$>$}}{\lower 0.511ex \hbox{$\sim$}}}}
\def\simle{\mathrel{
   \rlap{\raise 0.511ex \hbox{$<$}}{\lower 0.511ex \hbox{$\sim$}}}}
\def\vec#1{{\bm{#1}}}
\def\paragraph#1{}
\begin{document}
\title{Quantized Berry Phases of a Spin-1/2 Frustrated Two-Leg Ladder with Four-Spin Exchange
}
\author{I Maruyama$^1$, T. Hirano$^1$, and Y Hatsugai$^2$}
\address{
$^1$Department of Applied Physics, University of Tokyo, Hongo Bunkyo-ku, Tokyo 113-8656, Japan\\
$^{2}$Institute of Physics, Univ. of Tsukuba, 1-1-1 Tennodai, Tsukuba Ibaraki 305-8571, Japan
}
\ead{maru@pothos.t.u-tokyo.ac.jp}

\begin{abstract} 
A spin-1/2 frustrated two-leg ladder with four-spin exchange interaction
is studied by quantized Berry phases.
We found that
the Berry phase successfully characterizes
the Haldane phase
in addition to
the rung-singlet phase, and
the dominant vector-chirality phase.
The Hamiltonian of the Haldane phase
is topologically identical to the $S=1$ antiferromagnetic Heisenberg chain.
Decoupled models connected to the dominant vector-chirality phase
revealed that 
the local object identified by the non-trivial ($\pi$) Berry phase
is the direct product of two diagonal singlets.
\end{abstract}
\paragraph{introduction}

The $S=1/2$ two-leg ladder model with the four-spin exchange interactions
has been studied extensively
to clarify physics of La$_x$Ca$_{14-x}$Cu$_{24}$O$_{41}$\cite{PRB.62.8903,PRB.66.180404,PRL.98.027403}.
The four-spin ring exchange interaction
introduces frustration into the system
and
plays an essential role in several models
to give rise to exotic order, such as,  the vector chirality order\cite{PRL.95.137206},
nematic order\cite{PRL.96.027213}, and octapolar order\cite{PRL.97.257204}.
Especially,
phases in the two-leg ladder model with the four-spin ring exchange interactions
have been studied extensively
\cite{HikiharaDualPRL,JPSJ.77.014709,MomoiDualPRB,Laeuchli,FourspinEntangle,GritsevFourspin,SatoFourspin}.
To clarify its rich phases theoretically,
not only correlation functions corresponding to phases
but also
entanglement concurrence\cite{PRB.74.155119},
Lieb-Schulz-Mattis twist operators\cite{Laeuchli}, 
and quantized Berry phases\cite{AX.0806.4416}
have been studied.

The Berry phase
is pure quantum quantity due to quantum interference as in the case of the Aharonov-Bohm effect\cite{Berryphase}
and has no corresponding classical analogue.
The quantized Berry phase which is proposed to detect the topological and quantum orders
can be used as a order parameter even if there is no classical order parameter\cite{HatsugaiOrder1,HatsugaiOrder2,HatsugaiOrder3}.
It has been
successfully applied to gapped systems such as 
generalized valence bond solid states, dimerized Heisenberg models
\cite{HatsugaiOrderP,PRB.77.094431,HatsugaiOrder3}, 
surface states in semiconductors\cite{AX.0802.2425},
and the $t$-$J$ model\cite{Maruyama}.
An advantage of the Berry phase is that it quantizes to $0$
or $\pi$ even in the finite sized systems in any dimension
when the system has the time reversal invariance. 
Its quantization is protected against small perturbations
unless the gap closes.
This stability enables us to obtain
topologically identical models
which have the same result of Berry phases.
For these systems, 
the non-trivial ($\pi$) Berry phase reveals a local singlet or dimer,
which is a purely quantum object.

In the previous study\cite{AX.0806.4416},
the quantized Berry phases 
have been defined for models with four-spin exchange interactions,
and have been shown to be useful to characterize
the rung-singlet and dominant vector-chiral phases
in the $S=1/2$ spin ladder with ring exchange interactions.
It has also been shown 
that
the Hamiltonian of the rung-singlet phase 
is topologically identical to a decoupled rung-singled model
$H_{\rm RS}=\sum_{i=1}^{N/2}{\bf S}_{i,1}\cdot{\bf S}_{i,2}$
and 
that of the vector-chiral phase
is identical to a decoupled vector-chiral model
 $H_{\rm DVC} = \sum_{i=1}^{N/4} \left({\bf S}_{2i,1}\times{\bf S}_{2i,2}\right)\cdot
\left({\bf S}_{2i+1,1}\times{\bf S}_{2i+1,2}\right)$.
The latter Hamiltonian
includes only four-spin exchange interactions
and its ground state is a direct product of plaquette singlets.

In this paper,
we discuss other simple models 
which are topologically identical to $H_{\rm RS}$ or $H_{\rm DVC}$.
In addition, we shall extend the quantized Berry phase 
to the Haldane phase of the ladder.

Before detailed discussion of results,
let us describe the definition of the Berry phase
briefly.
Here,
the ladder model is generally written
as
$H=\sum_{i=1}^{N/2} 
\left[ J_r  h_{i,1;i,2}
+J_l \left( h_{i,1;i+1,1}+ h_{i,2;i+1,2}\right)
\right.$
$\left.
+J_d \left( h_{i,1;i+1,2}+ h_{i,2;i+1,1}\right)
+
J_{rr} h_{i,1;i,2} h_{i+1,1;i+1,2}
+
J_{ll} h_{i,1;i+1,1}  h_{i,2;i+1,2} 
+ 
J_{dd} h_{i,1;i+1,2}  h_{i,2;i+1,1} 
\right]
$
with $h_{i,\alpha;i',\alpha'} =\vec{S}_{i,\alpha} \cdot \vec{S}_{i',\alpha'}$,
where
all bilinear and quadratic terms in the Hamiltonian are
classified as rung, leg, and diagonal exchange interactions.
Especially We define a general frustrated ladder without four-spin exchange
interaction as 
$$H_{\alpha,\theta} = H(J_r=\cos\theta, J_l=\sin\theta, J_d = \alpha \sin \theta, J_{rr}= J_{ll}= J_{dd}=0).$$
The Berry phase $\gamma$ depends on how adiabatic parameter $\phi$ is introduced
to $H$.
For a parameter dependent Hamiltonian $H(\phi)$ with $H(\phi+2\pi)=H(\phi)$,
the Berry phase $\gamma$ is 
defined as $\gamma=\i \int_0^{2\pi}A(\phi)d\phi$ (mod $2\pi$), where $A(\phi)$ is the
Abelian Berry connection obtained by the single-valued
normalized ground state $|\mbox{gs}(\phi)\rangle$ of $H(\phi)$
as 
$A(\phi)=\langle\mbox{gs}(\phi)|\partial_\phi|\mbox{gs}(\phi)\rangle$.
The adiabatic parameter $\phi$ can be introduced by $h_{i,\alpha;i',\alpha'}(\phi)
={1\over 2}\left(
\e^{i\phi}S_{i,\alpha}^{+}S_{i',\alpha'}^{-}+ \e^{-i\phi}S_{i,\alpha}^{-}S_{i',\alpha'}^{+}
\right)+ 
S_{i,\alpha}^{z}S_{i',\alpha'}^{z}+S_{i,\alpha}^{z}S_{i',\alpha'}^{z}
$.
However,
we introduce the spin twist $\phi$ only on a specified link.
Then,
the Berry phase is used as a local order parameter defined on each link.
We consider three kinds of $\gamma$; the leg Berry phase $\gamma_l$,
the rung Berry phase $\gamma_r$, and
the diagonal Berry phase $\gamma_d$.
The rung-singlet phase is identified by $\gamma_r=\pi, \gamma_l=\gamma_d=0$
as realized in  $H_{RS}=H_{\alpha=0,\theta=0}$.
The dominant vector chirality phase is identified by $\gamma_d=\pi, \gamma_l=\gamma_r=0$.
If we introduce the spin twist $\phi$ on specified two links simultaneously,
we can define another Berry phase.

\paragraph{exact g.s.}
Under the condition $\alpha=1$ of $H_{\alpha,\theta}$,
there is a exact ground state\cite{IJMPB.12.2325}.
Using the operator $\vec{T}_i = \vec{S}_{i,1} + \vec{S}_{i,2}$,
the Hamiltonian is rewritten as
$H_{\alpha=1,\theta} = \cos\theta \sum_i \vec{S}_{i,1} \cdot \vec{S}_{i,2}
+\sin\theta \sum_i \vec{T}_i \cdot \vec{T}_{i+1}
= \cos\theta H_{RS}
+ \sin\theta H_{T}$.
There are two gapped phases:
the rung singlet phase and the Haldane phase.
The transition point is estimated around
$\theta_c \simeq 0.20\pi$\cite{PRB.52.12485,JPSJ.65.1387}.
It should be emphasized that
the ground state is unique and does not depend on $\theta$
in each phase.
Its ground state is 
either that of $H_{RS}$,
or that of $H_{T}$.
That is, the former is the rung singlet state
and the latter is the ground state of the $S=1$ antiferromagnetic Heisenberg chain.

\paragraph{Haldane twist}
The Berry phase of the $S=1$ antiferromagnetic Heisenberg chain
has been studied\cite{PRB.77.094431},
where 
the Berry phase has been defined through a $S=1$ spin twist to detect the Haldane phase.
That is,
a spin-flip term
$T_i^{+}T_j^{-}+T_i^{-}T_j^{+}$
is replaced with
$e^{i\phi}T_i^{+}T_j^{-}+e^{-i\phi}T_i^{-}T_j^{+}$,
where $\vec{T}$ is an operator of $S=1$.
It is natural to define the Berry phase of $H_{T}$ in the same manner
with replacing $\vec{T}_i = \vec{S}_{i,1} + \vec{S}_{i,2}$.
The corresponding spin twist
is 
$e^{i\phi}T_i^{+}T_j^{-}+e^{-i\phi}T_i^{-}T_j^{+} =
e^{i\phi}\left(S_{i1}^+ S_{i+1,1}^-
+S_{i2}^+ S_{i+1,2}^-
+S_{i1}^+ S_{i+1,2}^-
+S_{i2}^+ S_{i+1,1}^- \right)
+ h.c.$.
We define the Berry phase for the Haldane phase $\gamma_h$
by twisting all four bonds between $i$ and $i+1$ sites.
The fact that $\gamma_h=\pi$ 
indicates the Haldane phase of $H_{T}$ 
is explained by the valence bond solid picture
and confirmed by the previous study on the $S=1$ chain
\cite{PRB.77.094431}.

\begin{figure}
\begin{minipage}{5cm}
\resizebox{5cm}{5cm}{\includegraphics{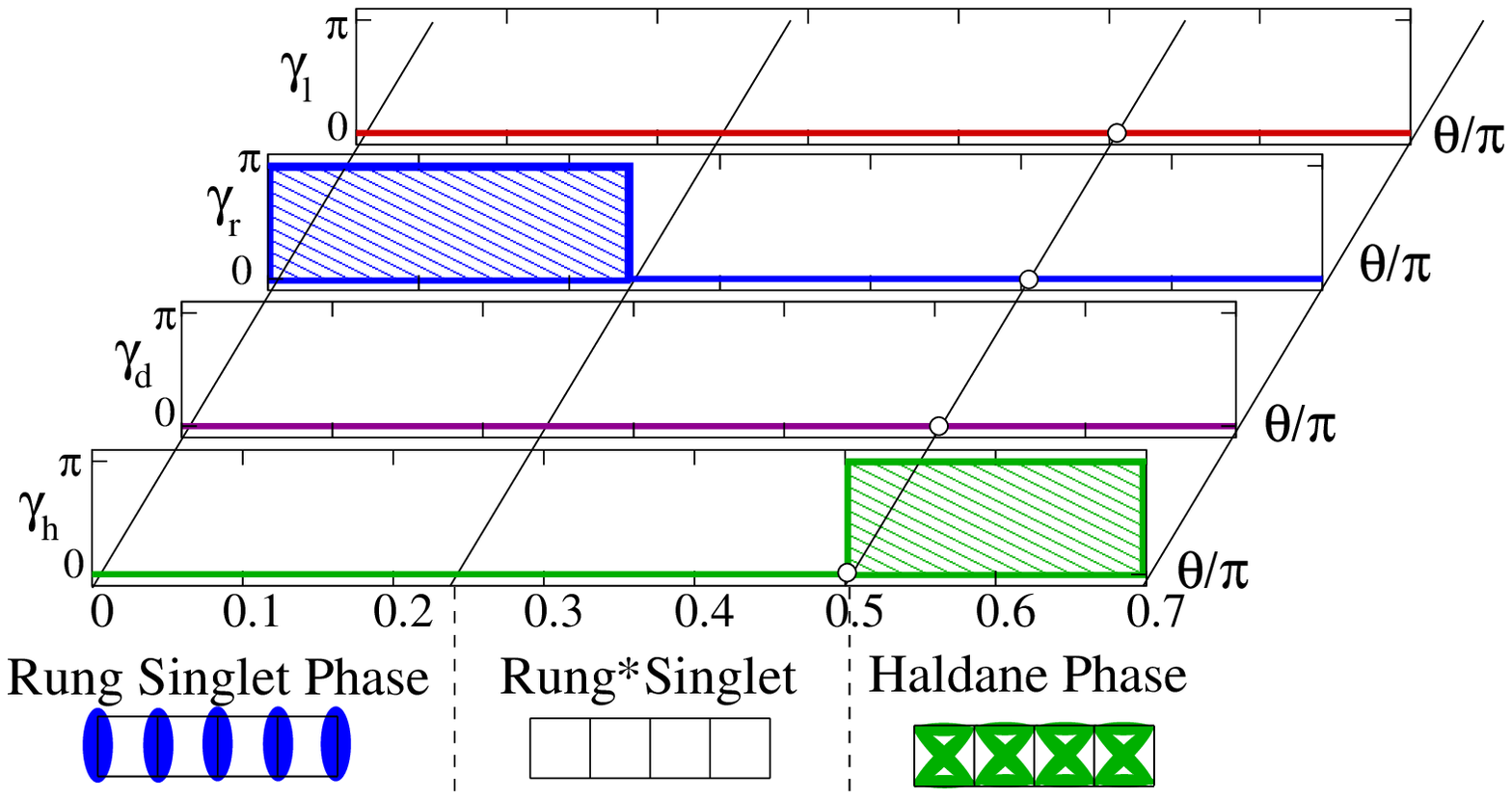}}
\caption{Berry phases as a function of $\theta$ of a ladder $H_{\theta,\alpha=0}$
at $N=16$.
}
\label{fig:Berrylad}
\end{minipage}
\hspace{1pc}%
\begin{minipage}{5cm}
\resizebox{5cm}{4.5cm}{\includegraphics{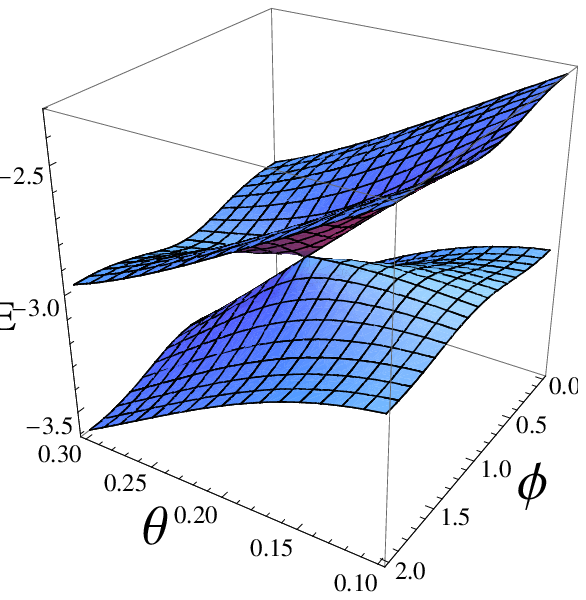}}
\caption{Energy diagram around gap-closing point between the rung singlet and rung$^*$ singlet phase
of $H_{\theta,\alpha=0}$
at $N=8$
with varying a spin twist $\phi$ on rung bond.
}
\label{fig:Ene}
\end{minipage}
\hspace{1pc}%
\begin{minipage}{5cm}
\resizebox{5cm}{5cm}{\includegraphics{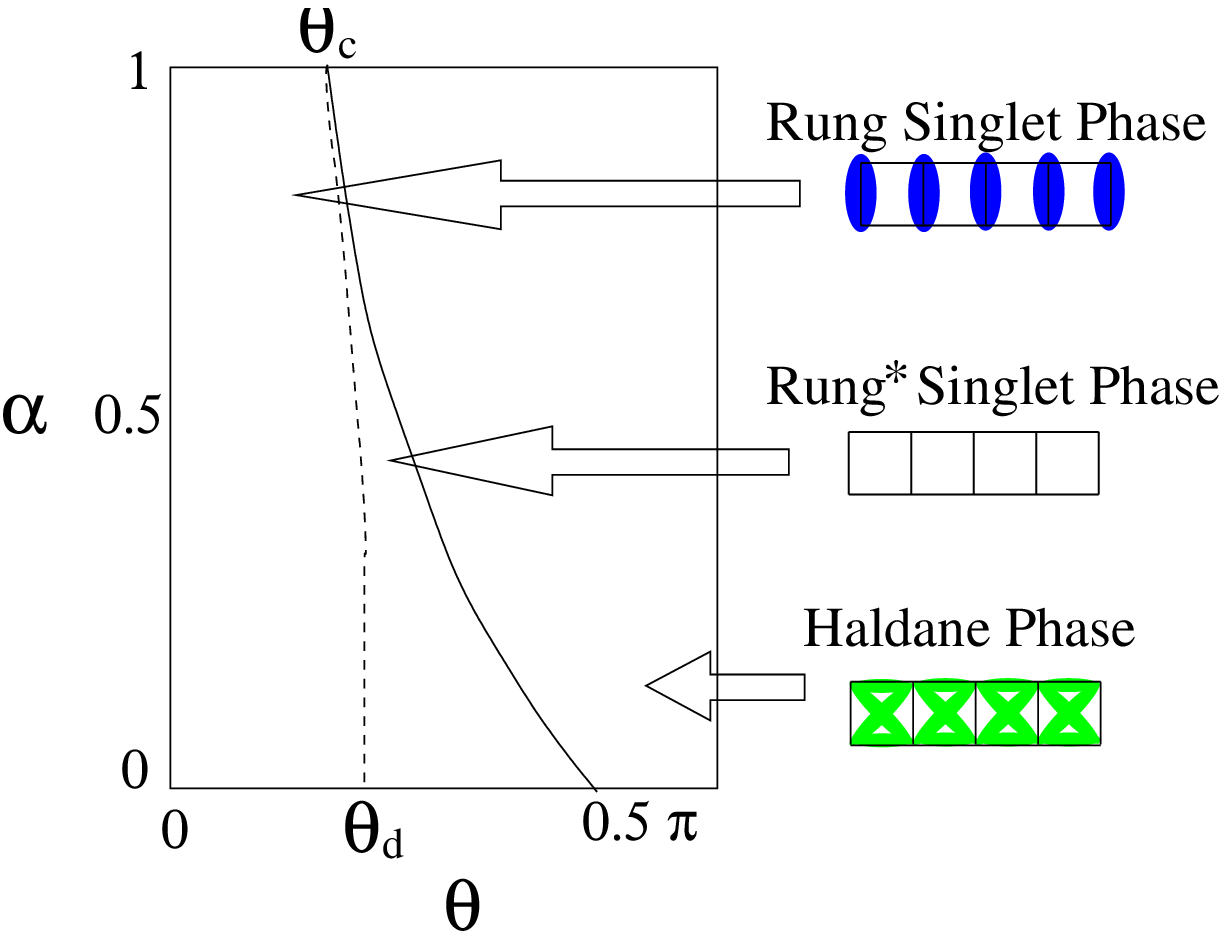}}
\caption{Schematic phase diagram of a frustrated ladder model $H_{\theta,\alpha}$
obtained at $N=8$.
}
\label{fig:phase}
\end{minipage}
\end{figure}

\paragraph{result}
In Ref.~\cite{AX.0806.4416},
the rung-singlet  phase
has been identified 
as $\gamma_r = \pi, \gamma_l=\gamma_d=0$ in a normal ladder $H_{\theta,\alpha=0}$.
In addition,
the rung$^*$ phase identified as $\gamma_r=\gamma_l=\gamma_d=0$
has been newly found.
It is naively expected that there be the Haldane phase when the rung bond is ferromagnetic
($\pi/2<\theta<\pi$).
To clarify the Haldane phase by $\gamma_h$ which has been introduced in this paper,
we calculated Berry phases of $H_{\theta,\alpha=0}$
for $0<\theta<\pi$.
Figure \ref{fig:Berrylad} shows the $\theta$ dependence of the
Berry phases at $N=16$
except for the gapless point $\theta=\pi/2$.
The Haldane phase is successfully identified by $\gamma_h$.
$\gamma_{d}=0$ is trivial because there is no diagonal interaction.

To discuss the transition between 
the rung-singlet and rung$^*$ phases,
the energy diagram as a function of $\theta$ and spin twist $\phi$
is shown in fig.~\ref{fig:Ene}.
Although 
the gap of untwisted Hamiltonian $H(\phi=0)$
is smoothly connected\cite{PRB.47.3196}
,
the gap of $H(\phi=\pi)$ for $\gamma_r$ 
closes at $\theta= \theta_d$.
Numerical data show $\theta_d < \pi/4$ and $\theta_d \sim 0.23 \pi$ at $N=20$.
The Berry phase $\gamma_r$ changes at $\theta_d$.
Then, the rung$^*$ phase has been identified by 
the quantum phase transition of $H(\phi=\pi)$,
which is not contradicting previous studies on the untwisted Hamiltonian ($\phi=0$),
which predict no transition between $0<\theta<\pi/4$.
However,
some papers
have implied
that the localized rung-singlet picture is limited to $0<\theta<\theta_d$ with $\theta_d<\pi/4$\cite{PRL.73.886,PRL.81.1941,PRB.47.3196,PRB.52.12485}.
Especially some new spin-liquid phase\cite{PRB.52.12485}
was predicted in a finite region in the phase diagram  of $H_{\alpha,\theta}$.
To clarify that it corresponds to the rung$^*$ phase,
we calculate the Berry phases of  $H_{\alpha,\theta}$.
The numerical result is summarized as the schematic phase diagram in fig.~\ref{fig:phase}.
The rung$^*$ phase existing in the normal ladder $H_{\theta,\alpha=0}$
disappears as $\alpha$ approaches to one.
At $\alpha=1$, there is a direct transition between the rung-singlet phase and the Haldane phase
at $\theta=\theta_c$.
$\theta_c = \mbox{Arctan} (J_l/J_r)_c \simeq 0.20\pi$ is determined by $(J_l/J_r)_c \simeq 0.71$.
This behavior is similar to that of the spin-liquid phase\cite{PRB.52.12485},
but the relation is still unclear.
Meaning of the rung$^*$ phase
can be revealed by different Berry phase associated with
twisting several links simultaneously as in the case of $\gamma_h$.

\paragraph{adiabatic continuation to decoupled models}
Finally,
we shall address
the decoupled vector-chiral model
$H_{\rm DVC}$.
Since $H_{\rm DVC}$ is decoupled,
it is enough to consider the 4 spin problem
$h_i^{(0)}$
,
where $h_i^{(0)}=\left({\bf S}_{2i,1}\times{\bf S}_{2i,2}\right)\cdot
\left({\bf S}_{2i+1,1}\times{\bf S}_{2i+1,2}\right)$
and $H_{\rm DVC}=\sum_i h_i^{(0)}$.
The Berry phase of $h_i^{(0)}$ is obtained as $\gamma_d=\pi,
\gamma_l=\gamma_r=0$.
After an adiabatic transformation,
we obtained three models $h_i^{(k)}$ with the same Berry phases.

\paragraph{ground state of HOBC}
The first model
is $h_{i}^{(1)}(\theta)$
with $J_r=J_l=\cos\theta+\sin\theta, J_d = \sin\theta$,
$J_{ll}=J_{rr}=4\sin\theta, J_{dd}=-4\sin\theta$,
which is the $N=4$ original Hamiltonian with the open boundary condition.
It can be easily shown that
$h_{i}^{(1)}(\theta)$
has $\gamma_d=\pi,
\gamma_l=\gamma_r=0$
between $\pi/2< \theta \simle 0.92\pi$.
The ground state of $h_{i}^{(1)}(\theta)$
does not depend on $\theta$
and is the direct product of two diagonal singlets.
For parameters $J_r=2 J_l, J_l=\cos\theta+\sin\theta, J_d = \sin\theta$,
$J_{ll}=J_{rr}=4\sin\theta, J_{dd}=-4\sin\theta$,
the ground state depends on $\theta$
while the Berry phases remain the same for $0.59 \pi\simle \theta \simle 0.93\pi$.

\paragraph{4 spin in}
The second model 
with the same Berry phases
is $h_{i}^{(2)}=- (\vec{S}_{2i,1} \cdot \vec{S}_{2i+1,2}) (\vec{S}_{2i,2} \cdot \vec{S}_{2i+1,1})$.
Its ground state is also  the direct product of diagonal singlets.
Moreover,
spin twist $\phi$ on a diagonal link
introduced in $h_{i}^{(2)}$
can 
gauged out
by the local gauge transformation of $\vec{S}_{2i,1}$ or $\vec{S}_{2i,2}$.
Since all spin is $S=1/2$,
it can be proved that $\gamma_d=\pi$.

\paragraph{diagonal dimer}
In addition, the third model with the same Berry phases
is a diagonal-singlet model $h_{i}^{(3)}=\vec{S}_{2i,1} \cdot \vec{S}_{2i+1,2}
+ \vec{S}_{2i,2} \cdot \vec{S}_{2i+1,1}$.
Its ground state is obviously the direct product of diagonal singlets.
This model reveals that non-trivial diagonal Berry phase $\gamma_d=\pi$
comes from local diagonal singlets,
since
$h_{i}^{(3)}$
includes only two-spin interactions.
It should be noted that 
$\gamma_d$ is defined by the spin twist on its diagonal link
which affects both two-spin and four-spin interactions
in an adiabatic transformation.
If the Berry phase is defined by the spin twist only in the two-spin interactions,
the Berry phase does not remains the same in the adiabatic transformation to the original Hamiltonian with four-spin ring exchange.

\paragraph{local object}
These three models $h_{i}^{(k)}, (k=1,2,3)$ are topologically identical to $h_{i}^{(k)}$
through the adiabatic transformation $(1-\alpha) h_{i}^{(0)} + \alpha h_{i}^{(k)}$
with adiabatic parameter $\alpha$, which has been confirmed numerically.
Each ground state of $h_{i}^{(k)}, (k=1,2,3)$ is a direct product of the diagonal singlets.
The spin twist $\phi$ introduced into $h_{i}^{(k)}, (k=2,3)$
can be gauged-out.
This fact leads to $\gamma_d=\pi$.
That is, a local object corresponding to $\gamma_d=\pi$ of $h_{i}^{(k)}, (k=2,3)$
is the direct product of two diagonal singlets
in the sense that  the local object for $\gamma_r=\pi$ of $H_{RS}$ is a localized rung singlet.
Note that the spin twist $\phi$ introduced into $H_{RS}$  is also able to be gauged-out.

\paragraph{local object}
In summary,
it has been shown that
the Berry phase for the Haldane phase $\gamma_h$
is useful for a frustrated two-leg ladder
as $\gamma_l, \gamma_r$, and $\gamma_d$ are useful in the previous study\cite{AX.0806.4416}.
Topological identification is established by adiabatic transformation
to the models with exact ground states.

\appendix
I.M. acknowledges discussions with S. Miyahara.
This work has been supported
in part by Grants-in-Aid for Scientific Research,
No. 20340098, 20654034 from JSPS and 
No. 220029004, 20046002 on Priority Areas from MEXT.
Some numerical
calculations were carried out on Altix3700BX2 at YITP in Kyoto
University
and the facilities of the Supercomputer Center, 
Institute for Solid State Physics, University of Tokyo.

\bibliography{../macro,../wiki,../book,manuscript}
\end{document}